\documentclass{article}

\usepackage{amsfonts}
\usepackage{amstext}
\usepackage{color}
\usepackage{amsthm}
\theoremstyle{plain}
\definecolor{Red}{cmyk}{0,1,1,0}

\definecolor{Blue}{cmyk}{1,1,0,0}

\definecolor{Pink}{cmyk}{0,1,0,0}

\definecolor{Green}{cmyk}{1,0,1,0.5}

\newcommand{\ba}{\begin{array}}
\newcommand{\ea}{\end{array}}
\newcommand{\be}{\begin{equation}}
\newcommand{\ee}{\end{equation}}
\newcommand{\ben}{\begin{enumerate}}
\newcommand{\een}{\end{enumerate}}

\newcommand{\eop}{\hfill \rule{0.7ex}{1.6ex}}

\newcommand{\R}{\mathbb{R}}

\newcommand{\B}{{\cal B}}

\newcommand{\N}{\mathbb{N}}

\newtheorem{theorem}{Theorem}[section]
\newtheorem{lema}{Lemma}[section]

\title{
{\large{ \bf{ Asymptotics for Nonlinear Integral Equations with a Generalized Heat Kernel using Renormalization Group Technique II:
Marginal Perturbations and Logarithmic Corrections to the Time Decay of Solutions}}}}

\begin{document}
\maketitle

\centerline{\scshape Gast\~ao A. Braga}
\medskip
{\footnotesize
 \centerline{Departamento de Matem\'atica}
  \centerline{Universidade Federal de Minas Gerais}
   \centerline{Caixa Postal 1621, Belo Horizonte, 30161-970, Brazil}
} 

\medskip

\centerline{\scshape Jussara M. Moreira}
\medskip
{\footnotesize
 \centerline{Departamento de Matem\'atica}
  \centerline{Universidade Federal de Minas Gerais}
   \centerline{Caixa Postal 1621, Belo Horizonte, 30161-970, Brazil}
}

\medskip

\centerline{\scshape Camila F. Souza}
\medskip
{\footnotesize
 \centerline{Departamento de Matem\'atica}
  \centerline{Centro Federal de Educa\c c\~ao Tecnol\'ogica de Minas Gerais}
   \centerline{Av. Amazonas, 5.253, Nova Sui\c ca, 30421-169, Brazil}
}

\def\l{\lambda}

\baselineskip = 22pt

\maketitle

\begin{abstract}
In this paper, we proceed with the analysis started in \cite{bib:braga-mor-souza}
and, using the Renormalization Group method, we obtain logarithmic corrections to the decay of
solutions for a class of nonlinear integral equations whenever the nonlinearities are
classified as marginal in the Renormalization Group sense.
\end{abstract}
\clearpage

\section{Introduction}
\label{sec:intr}

In this paper, we proceed with the analysis started in \cite{bib:braga-mor-souza}
where we employed the Renormalization Group (RG) method as developed by Bricmont et al.
\cite{bib:bric-kupa-lin} to obtain the long-time behavior of solutions to the
integral equation
$$
u(x,t)=\int_\R {G(x-y, s(t))f(y)dy}\,\, +
$$
\begin{equation}
\label{equ:nao:lin:int}
\int_1^{t}\int_\R {G(x-	y,s(t)-s(\tau))F(u(y,\tau))dy d\tau},~~ x\in\R\mbox{ and } t>1,
\end{equation}
where the integral kernel $G(x, t)$ satisfies the following three 
general conditions which we denote by {{\bf (G)}}:
\begin {enumerate}
\item[$(i)$]
There are integers $q >1$ and $M>0$ such that $G(\cdot,1)\in C^{q+1}(\R)$ and
$$
\sup_{x\in\R}\{(1+|x|)^{M+2}|G^{(j)}(x, 1)|\}<\infty, ~~~~j=0,1,..., q +1,
$$
where $G^{(j)}(x, 1)$ denotes the j-th derivative $(\partial_x^jG)(x, 1)$.
\item[$(ii)$]
There is a positive constant $d$ such that
$$
G(x,t)=t^{-\frac{1}{d}}G\left(t^{-\frac{1}{d}}x,1\right),~~~~ x\in\R,~t>0;
$$
\item [$(iii)$]
$
G(x,t)= \int_\R{G(x-y, t-s)G(y,s)dy}$, for $x\in\R$ and $t>s>0
$.
\end{enumerate}

In \cite{bib:braga-mor-souza} we considered the above conditions on $G(x,t)$ and 
nonlinearities $F(u)$ given by a power series of $u$
$F(u)=\lambda \sum_{j \geq \alpha}{a_ju^j}$,
where $\lambda \in [-1,1]$ and $\alpha$ is an integer satisfying $\alpha > \alpha_c$, with
\begin{equation}
\label{def:alph:crit}
\alpha_c = \frac{p+1+d}{p+1}.
\end{equation}
The parameter $d$ is given in $(ii)$ of $\textbf{(G)}$ and $p$ is assumed to be positive and it is
associated with the function $s(t)$ (see the argument of $G(x,t)$ in (\ref{equ:nao:lin:int})), which is such 
that $s(t)\sim t^{p+1}$ as $t\to \infty$. 
Furthermore, $f\in\B_q $, where $\B_q$ is the Banach space (\ref{def:Bq}). 
We have shown that, if $f$ is small in some sense, then 
the asymptotics of the solution to (\ref{equ:nao:lin:int}) is dictated by the
asymptotics of the integral kernel $G(x,t)$,
that is,
\begin{equation}
\label{def:irre-beha}
u(x,t) \sim \frac{A}{t^{(p+1)/d}}G\left(\frac{x}{t^{(p+1)/d}},
\frac{1}{p+1}\right)\mbox{ as } t\rightarrow \infty,
\end{equation}
where $A = A(p, f, \lambda, F)$.

The condition $\alpha>\alpha_c$ assumed in \cite{bib:braga-mor-souza} restricts the sum of $F$ 
to {\em irrelevant} (in the RG sense) perturbations. The main contribution
of this paper is to consider {\em marginal}  perturbations $u^{\alpha_c}$, besides irrelevant
ones. As we will see, marginal perturbations 
generate logarithmic corrections to the decay (\ref{def:irre-beha}). 
More specifically, we consider
\begin{equation}
\label{def:marg-pert}
F(u)= -\mu u^{\alpha_c} + \lambda\sum_{j \geq\alpha}{a_ju^j},
\end{equation}
where, in the above sum $\alpha$ is an integer satifying $\alpha > \alpha_c$, with $\alpha_c$ an integer given by (\ref{def:alph:crit}), and
we shall prove in Theorem \ref{teo:pri:cas:mar} that, if 
with  $\mu>0$ small and $\lambda\in \R$, then a logarithmic correction to the decay
(\ref{def:irre-beha}) shows up as follows
\begin{equation}
\label{def:marg-beha}
u(x,t) \sim \frac{A}{(t\ln t)^{(p+1)/d}}G\left(\frac{x}{t^{(p+1)/d}},
\frac{1}{p+1}\right)\mbox{  when } t\rightarrow \infty.
\end{equation}

For the nonlinear diffusion equation with time-dependent 
diffusion coefficient $c(t)=t+o(t)$
and with marginal perturbations, 
the long time behaviour (\ref{def:marg-beha}) with $d=2$, 
where $G(x,t)$ is 
to be replaced by the heat kernel, was obtained by Braga and
Moreira  in \cite{bib:Gas:Jus:1}. Here, we generalize their results 
to the integral equation (\ref{equ:nao:lin:int}) with 
nonlinearities $F(u)$ given by (\ref{def:marg-pert}) and 
integral kernels $G(x,t)$ satisfying
the condition $\textbf{(G)}$. We refer the reader to the Introduction
of \cite{bib:braga-mor-souza} for references on many interesting 
physical and engeneering problems which are modeled by equations 
with time-dependent difusions coefficients. 

To state our result, suppose $G(x,t)$ is given and 
condition $\textbf{(G)}$ is satisfied. Define
\begin{equation}
\label{def:Bq}
\B_q \equiv \{f:\R\rightarrow\R~|~\widehat f(\omega)\in C^1(\R)\mbox{ and }\|f\|<\infty\},
\end{equation}
where $\|f\|=\sup(1+|\omega|^q)(|\widehat f(\omega)|+|\widehat{f}'(\omega)|)$ and
$q>1$ is the integer given in $(i)$ of $\textbf{(G)}$.
 
We will consider (\ref{equ:nao:lin:int}) under the following hypothesis: 
\begin{enumerate}
\item[{\textbf{(M)}}]
\begin{enumerate}
\item[\,\,\ $(M_1)$] $F(u)$ is given by (\ref{def:marg-pert}), with $\mu>0$ and $\lambda\in\R$,
and the exponent $\alpha_c$, given by (\ref{def:alph:crit}), is assumed to be an integer, 
i.e., $p$ and $d$ are chosen so that $\alpha_c$ is an integer;
\item[\,\,\ $(M_2)$] Given $A_0>0$, let $g_0\in\B_q$ such that
$\widehat g_0(0)=0$ and $\|g_0\|<A_0^{\alpha_c}$. Consider initial 
conditions of form $f=A_0f^*_p+g_0$, with $f^*_p$ given by
\begin{equation}
\label{def:f_p^*}
f_p^*(x)=G\left(x, \frac{1}{p+1}\right);
\end{equation}
\item[\,\,\ $(M_3)$] 
the function $s(t)$, see the argument of
$G(x,t)$ in (\ref{equ:nao:lin:int}),  is given by $s(t)=\int_1^t c(\tau)d\tau$, 
with $c(t)$ a positive function in $L^1_{{loc}}(1, +\infty)$, 
given by $c(t)=t^p + o(t^p)$, with $p > 0$ and such that 
$$
\frac{1}{L^{n(p+1)}}\int_1^{L^n}{|o(t^p)|dt}\leq\frac{1}{n^{(p+1)/d}}, \,\,\, L>1, \,\,\, n\gg 1.
$$
\end{enumerate}
\end{enumerate} 
{\bf Remark:} Notice that, from hypothesis ($M_3$), 
\begin{equation}
\label{eq:s(t)}
s(t) = \frac{t^{p+1} -1}{p+1} + r(t),\,\,\, p>0,\,\,\, t\geq 1,
\end{equation}
where $r(t)$ satisfies
\begin{equation}
\label{eq:r(t)-bound}
\left|\frac{r(L^n)}{L^{n(p+1)}}\right|\leq \frac{1}{n^{(p+1)/d}}, 
\,\,\, L>1, \,\,\, n\gg 1.
\end{equation} 
The representation (\ref{eq:s(t)}) for $s(t)$ and the upper bound 
(\ref{eq:r(t)-bound}) for $r(t)$ are motivated by our previous experience, see
\cite{bib:Gas:Jus:1}, where we considered the special case $p=1$ and $d=2$. 
In there,  $s(t)=\int_1^t c(\tau)d\tau$, $c(t) \in L^1_{\it{loc}}(1, +\infty)$,
$c(t) = t + o(t)$ and $o(t)$ is a little order
of $t$ as $t\to\infty$ satisfying $L^{-2n}\int_1^{L^n} |o(t)|dt \leq n^{-1}$, $L>1$, $n\gg 1$.

The aim of this paper is to prove the following theorem:
\begin{theorem}
\label{teo:pri:cas:mar} Let $G(x,t)$ satisfying 
condition $\textbf{(G)}$ be given and consider equation 
(\ref{equ:nao:lin:int}) under the hypothesis $\textbf{(M)}$. There exist
positive constants $A$, $\lambda^*$ and $\epsilon$ such that\newline
$1)$ if $|A_0|<\epsilon$, where $A_0$ is given in ($M_2$);\newline
$2)$ if $ 0< \mu +|\lambda|<\lambda^*$, with $|\lambda| < \mu$,
where $\lambda$ and $\mu$ are given in ($M_1$),\newline
then the solution $u$ to the integral equation (\ref{equ:nao:lin:int}) satisfies
\begin{equation}
\label{eq:pri:cas:mar}
\lim_{t\rightarrow\infty}\|(t\ln t)^{(p+1)/d} u(t^{(p+1)/d}\cdot,t)-A f_p^*(\cdot)\|=0,
\end{equation}
where $f_p^*(\cdot)$ is given by (\ref{def:f_p^*}).
\end{theorem}

{\bf Remark:} The pre-factor $A$ multiplying $f_p^*(\cdot)$ in (\ref{eq:pri:cas:mar})
can be explicitly found and we will show that
\begin{equation}
\label{eq:pre-fact}
A =\left\{\left(\frac{d}{p+1}\right)\left[\frac{p+1}{(2\pi)^{d}}\right]^{\frac{1}{p+1}}\mu\,\, R\right\}^{-(p+1)/d}
\end{equation}
where $R$ depends upon $G(x,t)$ and is given by (\ref{def:R}).

This paper is a follow up of \cite{bib:braga-mor-souza}, being heavely based on it. 
In Section \ref{sec:critical} we quickly review the results in \cite{bib:braga-mor-souza}
which are important for this paper. In Section \ref{sec:reno} we prove the 
Renormalization Lemma and some other results which will be
important in Section \ref{sec:comp:assi:cas:mar} where we prove Theorem \ref{teo:pri:cas:mar}.
The heuristics behind the logaritmic correction to the decay is given in the Remark
right after the proof of Lemma \ref{estimativas}.
\section{\large{The RG operator}}
\label{sec:critical}

The RG approach consists in relating the long-time behavior 
of solutions to equations to the existence and stability of
fixed points of an appropriate RG transformation. By iterating the 
method, the RG transformation progressively evolves the solution 
in time and, simultaneously,
renormalizes the various terms of the equation under analysis. 
In \cite{bib:braga-mor-souza} 
we established the RG method to the integral equation (\ref{equ:nao:lin:int})
when $F(u)$ is irrelevant in the RG sense. 
With some adaptations, we will show that the method also works when we add marginal perturbations.
In order to study the nonlinear problem (\ref{equ:nao:lin:int}), with $F(u)$ in the form 
(\ref{def:marg-pert}), 
we recall some definitions and results from \cite{bib:braga-mor-souza} 
regarding the RG 
operator for the linear problem.

Given 
a time scale $L>1$ and a function $f$, define $f_0 \equiv f$, 
and 
\begin{equation}
\label{def:ufn}
u_{n}^0(x,t)\equiv\int{G\left(x-y,s_n(t)\right) f_n(y)}dy, \,\,\, t\in(1,L],
\end{equation}
where 
\begin{equation}
\label{def:s_n(t)}
s_n(t)=\frac{t^{p+1}-1}{p+1}+r_n(t), 
\end{equation}
with $r_n(t)= [r(L^nt)-r(L^n)]L^{-n(p+1)}$, where $r(t)$ is 
the remainder in (\ref{eq:s(t)}) satisfying the upper bound 
(\ref{eq:r(t)-bound}). Furthermore, for $n= 0, 1, 2, \cdots$,
\begin{equation}
\label{def:rg:lin}
f_{n+1}(\cdot) \equiv R^0_{L,n}f_n(\cdot) \equiv  L^{(p+1)/d}u_{n}^0(L^{(p+1)/d}\cdot,L).
\end{equation}

In Lemma II.4 of \cite{bib:braga-mor-souza} we have proved that 
there exists  a $p$-dependent constant $L_1>1$ such that
\begin{equation}
\label{cot:s_n(L)}
\frac{1}{6(p+1)}<\frac{s_n(L)}{L^{p+1}}<\frac{3}{2(p+1)},\,\,\, \forall\,\, n\geq 0, \,\,\, \forall\,\, L>L_1,
\end{equation}
and that there are positive constants 
$\tilde{K}$, $M$, $C_{d,p,q}$ depending on $d, p, q$ such that, 
for all $L>L_1$ given,
\begin{equation}
\label{lem:pon:fix}
\|f_p^*\|< C_{d,p,q}, \,\, \|R^0_{L^n}f^*_p\|\leq \tilde{K} {\mbox{ and }} 
\|R_{L^n}^0f^*_p-f^*_p\|\leq M\left|\frac{r(L^n)}{L^{n(p+1)}}\right|^{\frac{1}{d}},
 \end{equation}
where we have denoted $R^0_{L}\equiv R^0_{L,0}$, and 
where $f^*_p$ is defined by (\ref{def:f_p^*}). We have also proved 
the Contraction Lemma, see Lemma II.5 of \cite{bib:braga-mor-souza}, 
which asserts that there exists a 
constant $C=C(d,p,q)>0$ such that
\begin{equation}
\label{lema:contr}
\|R^0_{L,n}g\|\leq \frac{C}{L^{(p+1)/d}}\|g\|,~~\forall~L>L_1 {\mbox{ and }} n=0,1,2,\cdots,
\end{equation}
whenever $g\in \B_q$ is such that $\widehat g(0)=0$.

For the nonlinear equation (\ref{equ:nao:lin:int}), with
$F$ given by (\ref{def:marg-pert}), we fix $L>1$ and formally consider $\{u_n\}_{n=0}^{\infty}$ defined by
\begin{equation}
\label{def:un:cas:irr}
u_n(x,t)\equiv L^{n(p+1)/d}u(L^{n(p+1)/d}x,L^nt), ~~ t\in [1, L], ~~ n=0, 1, 2, \cdots,
\end{equation}
and with $u$ solution to 
\begin{eqnarray}
\label{equ:int:cas:mar}
u(x,t)&=&\nonumber \int {G(x-y, s(t))f(y)dy}-\mu\int_1^{t}\int {G(x-y,s(t)-s(\tau))u^{\alpha_c}(y,\tau)dy d\tau}\\
&+& \lambda  \int_1^{t}\int {G(x-y,s(t)-s(\tau)) \left[\sum_{j\geq\alpha}a_ju^j(y, \tau) \right] dy d\tau},
\,\,\, t>1.
\end{eqnarray}
We recall that, in the above sum, $\alpha$ is an integer satifying $\alpha > \alpha_c$, with $\alpha_c$ given by (\ref{def:alph:crit}).
The renormalized equation, in the marginal case, is, for each $n=0, 1, \cdots$,
\begin{eqnarray}
\label{equ:int:cas:mar:ren}
u_n(x,t)&=&\nonumber \int  {G(x-y, s_n(t))f_n(y)dy}-\mu\int_1^{t}\int {G(x-y,s_n(t)-s_n(\tau))u_n^{\alpha_c}(y,\tau)dy d\tau}\\
&+& \lambda_n\int_1^t\int G(x-y,s_n(t)-s_n(q))F_{L,n}(u_{n}(y,q))dydq
\end{eqnarray}
where $s_n(t)$ is given by (\ref{def:s_n(t)}) and
\begin{equation}
\label{def:Fln}
F_{L,n}(u_n)=\sum_{j\geq\alpha}a_jL^{n(\alpha-j)(p+1)/d}u_n^j,
\end{equation}
\begin{equation}
\label{def:lambdan}
\lambda_n= L^{-n(p+1)(\alpha-\alpha_c)/d}\lambda,
\end{equation}
and $f_n(x)\equiv L^{n(p+1)/d}u(L^{n(p+1)/d}x,L^n)$, 
$n= 1, 2, \cdots$, $f_0=f$, where $u$ is the solution to (\ref{equ:int:cas:mar}).

In Lemma III.1 of \cite{bib:braga-mor-souza} we have 
proved that, given $n\in\N$ and $L>1$, there exists $\epsilon_n>0$ 
such that, if $\|f_n\|<\epsilon_n$, then the integral equation 
(\ref{equ:int:cas:mar:ren}) has a unique solution in
$$
B_{f_n}\equiv\{u_n\in B^{(L)}:\|u_n-u_{n}^0\|\leq \|f_n\|\}
$$
where $u_{n}^0$ is the solution to (\ref{equ:int:cas:mar:ren}) with $\mu=\lambda_n=0$
(equivalently, given by (\ref{def:ufn}))  and
$$
B^{(L)}=\left\{u:\R\times[1,L]\rightarrow\R; u(\cdot,t) \in \B_q, ~\forall~t\in [1,L], 
\|u\|_L=\sup_{t\in[1,L]}\|u(\cdot,t)\|<\infty\right\}.
$$
In fact, Lemma III.1 of \cite{bib:braga-mor-souza} is valid for 
equation (\ref{equ:int:cas:mar:ren}) with nonlinearity and 
coupling constant given by (\ref{def:Fln}) and (\ref{def:lambdan}), 
respectively.

Since we are now interested in the effect of the marginal 
term in the asymptotics, we will rewrite the operator
$T_n:B^{(L)}\rightarrow B^{(L)}$, as $T_n(u_n)\equiv u_{n}^0 +V_n(u_n)$, 
where $V_n=-M_n + N_n$, $n=0, 1, 2, \cdots$,
\begin{equation}
\label{def:N_n}
M_n(u_n)(x,t)= \mu \int_1^{t}\int {G(x-y,s_n(t)-s_n(\tau))u_n^{\alpha_c}(y,\tau)dy d\tau}
\end{equation}
and
\begin{equation}
\label{def:Nn(u)}
N_n(u_n)(x,t)=\lambda_n\int_1^{t}\int {G(x-y,s_n(t)-s_n(\tau))F_{L,n}(u_n(y,\tau))dy d\tau}.
\end{equation}
Therefore, there exists $\epsilon_n$ such that, if $\|f_n\|<\epsilon_n$, 
$T_n$ has a unique fixed point which is the unique solution   
$u_n(x,t)$ for the renormalized integral equation (\ref{equ:int:cas:mar:ren}) 
for $t \in [1,L]$, which leads to the definition of the RG operator 
for the nonlinear equation
\begin{equation}
\label{def:R_{L,n}}
L^{(p+1)/d}u_{n}(L^{(p+1)/d}x,L)  \equiv (R_{L,n}f_n)(x) = f_{n+1}(x)
\end{equation}
for $n\geq 0$, where $f_0 = f$.

\section{Renormalization}
\label{sec:reno}

In this section we obtain the Renormalization Lemma for the marginal case. 
As in the irrelevant case treated in \cite{bib:braga-mor-souza}, we write
$f_{n}=A_{n}R^0_{L^{n}}f_p^*+g_{n}$ but we shall see that in this case the
sequence $(A_n)$ goes to zero as $n\to\infty$ and we have to keep track of this 
convergence in a certain way, which will be done
in the next Lemma. 

From now on, we denote 
$(e^{s_n(t)\mathcal{L}}f)(x,t)\equiv\int {G(x-y, s_n(t))f(y)dy}$.
Remember that, in  (\ref{equ:nao:lin:int}), the nonlinearity $F(u)$ 
is given by (\ref{def:marg-pert}), where $\alpha_c\geq 2$
is an integer, 
$\mu>0$ and $\lambda\in\R$. Notice that if $\mu < 0$
then solutions may blow up at finite time, see ~\cite{bib:fujita,bib:levine}.  
We shall prove that, 
in this case, the nonlinearity affects the asymptotic behavior, 
adding a logarithmic factor in the decay rate of convergence. 

Before stating and proving the Renormalization Lemma, we recall 
from \cite{bib:braga-mor-souza} (see Lemma II.1 of \cite{bib:braga-mor-souza} for details), that
$\widehat G(\omega,t)$, $(\omega,t)\in \R\times [1,\infty)$, as well as 
\begin{equation}
\label{ctes:KK1}
K \equiv \sup_{\omega\in\R}|\widehat G(\omega,1)|, \,\,\,\ K_1 \equiv \sup_{\omega\in\R}|\widehat G'(\omega,1)|,
\end{equation}
are well defined and we can rewrite condition $(ii)$ of $\textbf{(G)}$ in the Fourier space as
\begin{equation}
\label{cond:1.2:tra G}
\widehat G(\omega,t)=\widehat G(t^{\frac{1}{d}}\omega,1),~~~~\mbox{ for } t>0 \mbox{ and } \omega\in\R.
\end{equation}
Also, condition $(iii)$ of $\textbf{(G)}$ implies that
\begin{equation}
\label{cond:1.3:tra G}
\widehat G(\omega,t)=\widehat G(\omega,t-s)\widehat G(\omega, s) ~~ t>s>0 \mbox{ and } \omega\in\R.
\end{equation}

Finally, defining
\begin{equation}
\label{def:nuk}
\nu^*_n(x) =  \nu_n^*(x, L) \equiv \int_0^{L-1}{e^{[s_n(L)-s_n(L-\tau)]
\mathcal{L}}(e^{s_n(L-\tau)\mathcal{L}}R^0_{L^n}f_p^*)^{\alpha_c}d\tau}
\end{equation}
and
$\beta_n \equiv \widehat{\nu_n^*}(0)$, it is not hard to see that 
$\|\nu_n^*\|\leq \bar C$ for all $n$, with
\begin{equation}
\label{cte:barC}
\bar{C}=(L-1)\left(\frac{C_*}{2\pi}\right)^{\alpha_c-1}
\{2K+K_1[3L^{p+1}/2(p+1)]^{1/d}\}^{\alpha_c+1}{\tilde{K}}^{\alpha_c},
\end{equation}
with $C_*\equiv (2^{q+1}+3)\int_{\R}{[1+|x|^q]^{-1}dx}$, $K$ and $K_1$ 
given in (\ref{ctes:KK1}) and $\tilde K$ the constant in (\ref{lem:pon:fix}).

\begin{lema}[Renormalization Lemma]
\label{estimativas}
Given $k\in \N$ and $L>L_1$, suppose that $f_n$ given by (\ref{def:R_{L,n}}) is well defined for $n=1,2,\cdots,k+1$.
Then, for each $n$, there is a constant $A_n$ and a function $g_n \in \B_q$ with $\widehat g_n(0)=0$ such that
\begin{equation}
\label{rep:fn}
f_0=A_0f_p^*+g_0,~~~~~~
f_{n+1}=A_{n+1}R^0_{L^{n+1}}f_p^*+g_{n+1}~~~~~~~(n=0,1,...,k).
\end{equation}
Furthermore, there exist n-independent positive constants $\gamma$ and $\Lambda$
such that, if $|A_n|\leq 1$,
$\|g_n\|\leq 1$ and $0<\mu+|\lambda|<\gamma$, then
$$
|A_{n+1}-A_n+\mu \beta_n A_n^{\alpha_c}|\leq \Lambda[\mu(|A_n|^{2\alpha_c-1}+ |A_n|^{\alpha_c-1}\|g_n\|+\|g_n\|^{\alpha_c})+
$$
\begin{equation}
\label{cotaA:m}
(\mu+1)|\lambda|(|A_n|^{\alpha_c+1}+\|g_n\|^{\alpha_c+1})].
\end{equation}
and
$$
	\|g_{n+1}\|\leq \frac{C}{L^{(p+1)/d}}\|g_n\|+\Lambda
	[\mu(|A_n|^{\alpha_c}+|A_n|^{2\alpha_c-1}+ |A_n|^{\alpha_c-1}\|g_n\|+\|g_n\|^{\alpha_c})+
	$$
	\begin{equation}
	\label{cotag}
	+(\mu+1)|\lambda|(|A_n|^{\alpha_c+1}+\|g_n\|^{\alpha_c+1})].
		\end{equation}
\end{lema}

{\bf{Proof: }} Decomposition (\ref{rep:fn}) follows from 
induction, exactly like in the proof of Lemma III.2 
of \cite{bib:braga-mor-souza}, where we have defined
for $n\geq 0$,
\begin{equation}
\label{def:An}
A_{n+1}=A_n+\widehat\nu_n(0)
\end{equation}
and
\begin{equation}
\label{def:g_j+1:cas:mar}
g_{n+1}(x)= R^0_{L,n+1}g_n(x)+L^{(p+1)/d}\nu_n(L^{(p+1)/d} x)-\widehat\nu_n(0)R^0_{L^{n+1}}f_p^*(x),
\end{equation}
with the difference that now $\nu_n(x)=V_n(u_n)(x,L) = -M_n(u_n)(x,L) + N_n(u_n)(x,L)$,
where $M_n(u_n)$ and $N_n(u_n)$ were defined in (\ref{def:N_n}) and (\ref{def:Nn(u)}), respectively.
In order to obtain estimates (\ref{cotaA:m}) and (\ref{cotag}), 
we define $w_n=\nu_n+\mu A_n^{\alpha_c}\nu_n^*$ with $\nu_n^*,$ 
given by (\ref{def:nuk}) and, since $\beta_n\equiv \widehat{\nu_n^*}(0)$, 
we have $\widehat w_n(0)=A_{n+1}-A_n+\mu A_n^{\alpha_c}\beta_n$. 
In Lemma \ref{Lem:est} we will prove that there exist positive 
constants $\gamma$ and $E$ such that, if $|A_n|\leq 1$, 
$\|g_n\|\leq 1$ and $0<\mu+|\lambda|<\gamma$, then
\begin{eqnarray}
\label{cot:wn}
 \|w_n\| \nonumber\leq E[\mu(|A_n|^{2\alpha_c-1} &+& |A_n|^{\alpha_c-1}\|g_n\|+\|g_n\|^{\alpha_c})\\
&+& (\mu+1)|\lambda|(|A_n|^{\alpha_c+1}+\|g_n\|^{\alpha_c+1})],
\end{eqnarray}
which will prove (\ref{cotaA:m}), for all $\Lambda \geq E$. 
In order to prove (\ref{cotag}), we use definition
(\ref{def:g_j+1:cas:mar}) and inequalities (\ref{lem:pon:fix}) and (\ref{lema:contr}) to obtain:
$$
\|g_{n+1}\|\leq \frac{C}{L^{(p+1)/d}}\|g_n\|+(L^{q(p+1)/d}+\tilde{K})\|\nu_n\|.
$$
Furthermore, since $\|\nu_n\|\leq \mu |A_n|^{\alpha_c}\|\nu_n^*\|+\|w_n\|$, we bound $\|g_{n+1}\|$ by
$$
\frac{C}{L^{(p+1)/d}}\|g_n\|+(\bar C+E)(L^{q(p+1)/d}+\tilde{K})
[\mu(|A_n|^{\alpha_c}+|A_n|^{2\alpha_c-1}+ |A_n|^{\alpha_c-1}\|g_n\|+\|g_n\|^{\alpha_c})+$$
$$
+(\mu+1)|\lambda|(|A_n|^{\alpha_c+1}+\|g_n\|^{\alpha_c+1})].
$$
Since $L>1$, defining
$\Lambda \equiv (\bar C+E)(L^{q(p+1)/d}+\tilde{K})$,
the proof is finished. 

\eop

{\bf Remark:} At this point it is possible to understand, heuristically,
how the logarithmic correction to the decay pops up and 
inequality (\ref{cotaA:m}) is crucial for that.  
In Lemma \ref{lem:4} we will show that
$\|g_n\|<A_n^{2}$ for all $n$ and that $A_n\rightarrow 0$ as 
$n\rightarrow\infty$. Together with $\alpha_c\geq 2$,  
this implies that the right hand side of (\ref{cotaA:m}) is a little order of 
$A_n^{\alpha_c}$, meaning that it can be dropped off when compared with
the left hand side so that
$$
A_{n+1}-A_n+\mu \left( R \left[\frac{p+1}{(2\pi)^{d}}\right]^\frac{1}{p+1} \ln L \right) A_n^{\alpha_c}\approx 0,
$$
where we have used that $\beta_n \approx  (R \left[\frac{p+1}{(2\pi)^{d}}\right]^\frac{1}{p+1} \ln L)$
as $n\to\infty$ (see Lemma \ref{lem:cond:beta}).
Integrating out the above equation gives
$$
A_n \approx A \left[
\frac{1}{\ln t_n}\right]^{\frac{p+1}{d}}, \,\,\,\,\,\, t_n = L^n,
$$
where $A$ is given by (\ref{eq:pre-fact}). For the rigorous argument, see the proof 
of Lemma \ref{lema:princ}, in particular Equation (\ref{cota:An:cas:marginal}).

From now on we will denote $u_{A_n}$ instead of $u_n$ to 
emphasize the relation between the solution and the 
decomposition of the initial data given by the Renormalization Lemma, 
that is, given $ L> L_1 $, let $u_{A_n}$ be the solution to 
(\ref{equ:int:cas:mar:ren}) with initial data
$f_n = A_nR^0_{L^n}f_p^* + g_n$. Furthermore, let $u^*_{A_n}$ 
be the solution to problem (\ref{equ:int:cas:mar:ren}) with 
$\lambda_n=0$ and initial data $f_n^* = A_nR^0_{L^n} f_p^*$. 
Notice that $u^*_{A_n}$ ``measures'' the effect of the critical nonlinearity
on the component of the initial condition which is in the direction of the
asymptotic fixed point of the linear RG operator. Therefore, if the norm of $g_n$ is small,
we expect that $ u_{A_n} $ is somehow ``close'' to $u_{A_n}^*$, 
which motivates the estimates we will obtain next. 
Notice that, for $w_n=\nu_n+\mu A_n^{\alpha_c}\nu_n^*$, with $\nu_n^*$ 
given by (\ref{def:nuk}), we can write down the upper bound
$$
		\|w_n\|\leq \|M_n(u^*_{A_n})(L)-\mu A_n^{\alpha_c}\nu_n^*\|
		+\|M_n(u_{A_n})-M_n(u^*_{A_n})\|_L+\|N_n(u_{A_n})\|_L.
$$
We will then obtain, in the next lemma, 
estimates for the norms on the right hand side of the
inequality above, thus proving (\ref{cot:wn}). 
We refer of $\epsilon_n$ given by (43) of \cite{bib:braga-mor-souza}.
\begin{lema}
\label{Lem:est}
Given $L>L_1$ and $n\in\{0,1,2,...\}$ suppose that the initial 
condition $f_n$ for problem (\ref{equ:int:cas:mar:ren})
can be written as $f_n=A_nR^0_{L^n}f^*_p+g_n$, with $g_n\in\B_q$, 
$\|g_n\|\leq 1$, $|A_n|\leq 1$ and $\|f_n\|<\epsilon_n$. 
Then, there exist positive constants $E$ and $\gamma$ such that, 
if $0<|\lambda|+\mu <\gamma$, then 
\begin{equation}
\label{equ:1:lem}
\|M_n(u_{A_n})-M_n(u_{A_n}^*)\|_L\leq \mu E[|A_n|^{\alpha_c-1}\|g_n\|
+\|g_n\|^{\alpha_c}+|\lambda|(|A_n|^{\alpha_c+1}+\|g_n\|^{\alpha_c+1})],
\end{equation}
\begin{equation}
\label{equ:2:lem}
\|M_n(u_{A_n}^*)(L)-\mu A_n^{\alpha_c}\nu_n^*\|_L\leq\mu E|A_n|^{2\alpha_c-1}
\end{equation}
and
\begin{equation}
\label{equ:3:lem}
\|N_n(u_{A_n})\|_L\leq |\lambda| E(|A_n|^{\alpha_c+1}+\|g_n\|^{\alpha_c+1}).
\end{equation}

\end{lema}
{\bf{Proof: }} First of all, since $\|f_n\|<\epsilon_n$, then $u_{A_n}$ and $u^*_{A_n}$ are the only solutions to the respective equations in $B_{f_n}$ e $B_{f_n^*}$ given by
\begin{equation}
\label{def:u*}
u^*_{A_n}(t)=A_ne^{s_n(t)\mathcal{L}}R^0_{L^n}f^*_p- M_n(u_{A_n}^*)(t)
\end{equation}
and
\begin{equation}
\label{def:u}
u_{A_n}(t)=A_ne^{s_n(t)\mathcal{L}}R^0_{L^n}f^*_p+e^{s_n(t)\mathcal{L}}g_n+V_n(u_{A_n})(t).
\end{equation}
Defining $\bar C_0 \equiv 2K+K_1[3L^{p+1}/2(p+1)]^{1/d}$, using (\ref{cot:s_n(L)})
and the properties of the kernel $G$, we get
\begin{equation}
\label{cota:caso:marg}
\|M_n(u_{A_n})-M_n(u_{A_n}^*)\|_L\leq \mu \alpha_c \bar{C_{1}}\|u_{A_n}-u_{A_n}^*\|_L(\|u_{A_n}\|_L^{\alpha_c-1}+\|u_{A_n}^*\|_L^{\alpha_c-1}),
\end{equation}
with $\bar{C_1}=\bar C_0(L-1)(C_*/2\pi)^{\alpha_c-1}$. 
Now we recall that, if $\|f_n\|<\epsilon_n$, then $\|u_{A_n}\|_L<\rho_0$ 
(see proof of Lemma III.1 in \cite{bib:braga-mor-souza}). Therefore, defining 
$S_1(z)=(C_*/2\pi)^{\alpha_c-1}\rho_0^{\alpha_c-1}+\sum_{j\geq \alpha}(C_*/2\pi)^{j-1}|a_{j}|z^{j-1}$, 
$\gamma_{0}=[2\bar C_0(L-1)S_1(\rho_0)]^{-1}$ and $\bar{C_2}=2(\tilde{K}+1)\bar C_0$,
taking $\mu+|\lambda|<\gamma_{0}$ gives
\begin{equation}
\label{cot:u*}
\|u^*_{A_n}\|_L\leq \bar{C_2}|A_n|,
\end{equation}
\begin{equation}
\label{cot:u}
\|u_{A_n}\|_L\leq\bar{C_2}(|A_n|+\|g_n\|)
\end{equation}
and therefore,
\begin{equation}
\label{cota:uAn}
\|u_{A_n}\|_L^{\alpha_c-1}+\|u_{A_n}^*\|_L^{\alpha_c-1}\leq 2\bar{C_2}^{\alpha_c-1}(|A_n|+\|g_n\|)^{\alpha_c-1}.
\end{equation}
Similarly, since $|\lambda_n|<|\lambda|$ for all $n$, 
\begin{equation}
\label{desig:Nn}
\|N_n(u_{A_n})\|_L\leq |\lambda|\bar{C_3}\bar{C_2}^{2}(|A_n|+\|g_n\|)^{2},
\end{equation}
with $\bar{C_3}=\bar C_0(L-1)S_2(\rho_0)$, where $S_2(z)=\sum_{j\geq\alpha}(C_*/2\pi)^{j-1}|a_j|z^{j-2}$.
Defining
$$
\gamma=\min\left\{1,\gamma_{0},\frac{1}{2^{\alpha_c+1}\alpha_c\bar{C_1} \bar{C_2}^{\alpha_c-1}}\right\},
$$
if $\mu<\gamma$, since $\|g_n\|\leq 1$ and $|A_n|\leq 1$, using (\ref{cota:uAn}) and (\ref{desig:Nn}) in (\ref{cota:caso:marg}), 
we get (\ref{equ:1:lem}) with $E=E_1\equiv 4(\alpha_c+2)!\alpha_c\bar{C_1} \bar{C_2}^{\alpha_c-1}(1+\bar{C_3}\bar{C_2}^{2})\left\{1+K+K_1[3L^{p+1}/2(p+1)]^{1/d}\right\}$.

In order to prove (\ref{equ:2:lem}), notice that we can write $M_n(u^*_{A_n})(L)=
\mu A_n^{\alpha_c}\nu_n^*+\mu\sum_{j=0}^{\alpha_c-1}I_j$, with $\nu_n^*$ given by (\ref{def:nuk}) and
$I_j$ given by
$$\left(
            \begin{array}{c}
              \alpha_c \\
              j \\
            \end{array}
          \right)
\int^{L-1}_0{e^{[s_n(L)-s_n(L-\tau)]\mathcal{L}}[(A_ne^{s_n(L-\tau)\mathcal{L}}R^0_{L^n}f^*_p)^j
[-M_n(u^*_{A_n})(L-\tau)]^{\alpha_c-j}]d\tau}.
$$
Noticing that $\|M_n(u^*_{A_n})\|_L\leq \mu \bar C_1(\bar C_2|A_n|)^{\alpha_c}$, if 
$C^*_j=\alpha_c!\tilde{K}^j\bar C_0^j\bar{C_1}^{\alpha_c-j+1}\bar{C_2}^{\alpha_c(\alpha_c-j)}$, 
$
\|I_j\|\leq \bar{C}^*_{j}|A_n|^{j+\alpha_c(\alpha_c-j)}\mu^{\alpha_c-j}.
$
Therefore, 
$$
\|M_n(u^*_{A_n})(L)-\mu A_n^{\alpha_c}\nu_n^*\| \leq \mu|A_n|^{2\alpha_c-1}\sum_{j=0}^{\alpha_c-1}\bar{C}_{j}^*|A_n|^{\alpha_c^2-\alpha_c(j+2)+j+1}\mu^{\alpha_c-j}
$$
and using that $|A_n|\leq 1$ and $\mu \leq 1$
we prove (\ref{equ:2:lem}) with $E=E_2\equiv \sum_{j=0}^{\alpha_c-1}\bar{C}_{j}^*$.

Finally, from (\ref{cot:u}) and from the fact that 
$\|N_n(u_{A_n})\|_L\leq|\lambda|\bar C_0S_3(\rho_0)\|u_{A_n}\|_L^{\alpha_c+1}$, where
$S_3(z)=\sum_{j\geq\alpha}(C_*/2\pi)^{j-1}|a_j|z^{j-\alpha_c-1}$, we obtain inequality (\ref{equ:3:lem}) with $E=E_3\equiv(\alpha_c+2)!\bar C_0S_3(\rho_0)\bar{C_2}^{\alpha_c+1}$. 
Defining $E\equiv \max\{E_1,E_2,E_3\}$, we conclude the proof.
\eop

\section{Asymptotic Behavior}
\label{sec:comp:assi:cas:mar}

In order to obtain the asymptotic limit (\ref{eq:pri:cas:mar}),
we first prove that $\beta_n(L) = \widehat{\nu_n^*}(0, L)$, $n = 0, 1, 2, \cdots$, 
where $\widehat{\nu_n^*}$ is given by (\ref{def:nuk}),
is a convergent sequence as $n\to\infty$.
Notice that, from the properties of the kernel $G$, the integral
\begin{equation}
\label{def:R}
R=\int\widehat G(-x_1,1)\widehat G(x_1-x_2,1) \cdots\widehat G(x_{\alpha_c-1},1)dx_1 \cdots dx_{\alpha_c-1}
\end{equation}
is well defined. 
\begin{lema}
\label{lem:cond:beta}
Consider Equation (\ref{equ:int:cas:mar}) under the hypothesis 
$\textbf{(M)}$, with $G(x,t)$ satisfying $\textbf{(G)}$. 
Let $\beta_n = \widehat{\nu_n^*}(0)$, 
with $\nu^*_n$ given by (\ref{def:nuk}) and 
\begin{equation}
\label{def:beta}
\beta=R \left[\frac{p+1}{(2\pi)^{d}}\right]^\frac{1}{p+1} \ln L,
\end{equation}
where $R$ is given by (\ref{def:R}).
Then, there exists a constant $C(d,L,p)$ such that
\begin{equation}
\label{conver:beta}
|\beta_n-\beta|\leq C(d,L,p)\left(\frac{1}{n}\right)^{\frac{p+1}{d}},
\end{equation}
\end{lema}
for $n$ sufficiently large. 

{\bf{Proof: }} In what follows, we drop off the $L$-dependence on functions and parameters.
Defining $g(y,\tau)\equiv \int{G(y-z, s_n(L-\tau))R^0_{L^n}f_p^*(z)dz}$ and observing that
$\hat G(0,t)=1$ for $t>0$, we get from (\ref{def:nuk}) that 
\begin{equation}
\label{equ:betak}
\beta_n=\int_0^{L-1}{\widehat{\left[g^{\alpha_c}(\cdot,\tau)\right]}|_{\omega = 0}\,\, d\tau}.
\end{equation}
Using 
the definition of the RG operator and properties of $G$, from 
(\ref{cond:1.2:tra G}) and (\ref{cond:1.3:tra G}) we get
$$
\hat g(\omega,\tau)=\hat G\left(\omega,s_n(L-\tau)+\frac{1}{L^{n(p+1)}}\left[s(L^n)+\frac{1}{p+1}\right]\right)
$$
so that, from above and from properties of the Fourier Transform, we can rewrite $\beta_n$ in (\ref{equ:betak}) as
$$
\frac{1}{(2\pi)^{\alpha_c-1}}\left[ \int_0^{L-1} 
\int_ {\R^{\alpha_c-1}}\hat G(-p_1,a)
\hat G(p_1-p_2,a)\cdots \hat G(p_{\alpha_c-1},a)dp_1dp_2\cdots dp_{\alpha_c-1}\right] d\tau,
$$
where $a=s_n(L-\tau)+L^{-n(p+1)}\left[s(L^n)+1/(p+1)\right]$. 
Now, recalling that $\alpha_c = (p+1+d)/(p+1)$, using definitions (\ref{def:R}) and (\ref{def:s_n(t)}) of $R$ and $s_n(t)$, we get
\begin{equation}
\label{betak}
\beta_n=R \left[
\frac{(p+1)^{1/p+1}}{(2\pi)^{\alpha_c-1}}\right]
\int_0^{L-1}\left[(L-\tau)^{p+1}+h_n\right]^{-1/(p+1)}d\tau,
\end{equation}
where $h_n=(p+1)[r_n(L-\tau)+L^{-n(p+1)}r(L^n)]$. 
Noticing that $h_n\rightarrow 0$ when $n\rightarrow\infty$, we conclude
that $\beta_n \to \beta$ converges as $n\to\infty$. 
Furthermore, 
\begin{equation}
\label{dif:beta}
|\beta_n-\beta|= |R| \left[
\frac{(p+1)^{\frac{1}{p+1}}}{(2\pi)^{\alpha_c-1}}\right]
\left|\int^{L-1}_0\int_0^{h_n} \frac{1}{(p+1)[(L-\tau)^{p+1}+\bar h]^{\frac{p+2}{p+1}}}d\bar h
d\tau\right|.
\end{equation}
Taking $n$ sufficiently large so that $\bar h>-\frac{3}{4}$ and using the definition of $h_n$, 
$$
|\beta_n-\beta|\leq 
S(d,p)\left[\int^{L}_1{\frac{|r_n(t)|}{t^{p+2}}dt}+
\left|\frac{r(L^n)}{L^{n(p+1)}}\right|\int^{L}_1{\frac{1}{t^{p+2}}dt}\right],
$$
where $S(d,p)=|R|[4^{p+2}(p+1)]^{\frac{1}{p+1}}/(2\pi)^{\alpha_c-1}$. 
Using condition $(M_3)$ of $\textbf{(M)}$ in the definition of $r_n$, we have that
$|r_n(L)|\leq L^{p+1} (n+1)^{-(p+1)/d}$ and therefore
$$
\left|\frac{r(L^n)}{L^{n(p+1)}}\right|\leq
\left(\frac{1}{n}\right)^{\frac{p+1}{d}},
$$
which leads to (\ref{conver:beta}), with $C(d,L,p)=S(d,p)[L^{2(p+1)}-1](p+1)^{-1}L^{-(p+1)}$.
\eop

We notice that, if $L>L_1$, since $s_n(t)$ is an increasing 
function for all $n\geq 0$ and $0\leq\tau\leq L-1$, it follows that
$$
\left(\frac{1}{6(p+1)}\right)^{\frac{1}{p+1}}<\left[s_n(L-\tau)+\frac{1}{L^{n(p+1)}}\left(s(L^n)+
\frac{1}{p+1}\right)\right]^{\frac{1}{p+1}}<L\left(\frac{4}{p+1}\right)^{\frac{1}{p+1}}
$$
and therefore $\beta_*<\beta_n<\beta^*$ for all $n\geq 0$,
where
$$
\beta_*= \frac{R}{(2\pi)^{\alpha_c-1}}
\left[
\frac{p+1}{4}\right]^{\frac{1}{p+1}}\left[1-\frac{1}{3^{1/(p+1)}}
\right]  
{\mbox{ and }} \beta^*=\frac{R}{(2\pi)^{\alpha_c-1}}(L-1)[6(p+1)]^{\frac{1}{p+1}}.
$$
We will use the previous bounds in the next lemma, where we 
prove that $(A_n)$ is a decreasing sequence that goes to zero 
when $n$ goes to infinity, which will allow us to obtain the 
unique global solution to the problem. In Lemma \ref{lem:4} 
we make use of the definition $L_2\equiv \max\{L_1,C^{d/(p+1)}\}$ 
introduced in the proof of Theorem II.1 of 
\cite{bib:braga-mor-souza} and we refer to $\sigma$ 
given by (45) in \cite{bib:braga-mor-souza}, which is a lower bound for the sequence
$(\epsilon_n)$.

\begin{lema}
\label{lem:4}
For $L>L_2$, there are positive constants $\epsilon$ and $\lambda^*$ such that, if $0<\mu+|\lambda|<\lambda^*$,
$|\lambda|<\mu$ and $f_0=A_0f_p^*+g_0$ with $A_0\in(0,\epsilon)$, $\|g_0\|<A_0^2$ and $\hat g_0(0)=0$, then $f_{n+1}=R_{L,n}f_n$ is well defined for $n=0,1,2,\dots$ and (\ref{rep:fn}) is valid with $A_{n+1}$ and $g_{n+1}$ given by (\ref{def:An}) and (\ref{def:g_j+1:cas:mar}), respectively. Furthermore,
$0<A_{n+1}<A_n$, $\|g_n\|<A_n^{2}$ for all $n$ and $A_n\rightarrow 0$ when $n\rightarrow\infty$.
\end{lema}
\textbf{Proof: } Notice that if $\epsilon\leq 1$, since
$A_0<\epsilon$ and $\|g_0\|<A_0^2$, we get $A_0<1$ and $\|g_0\|<1$. 
Furthermore, since $f_0=A_0f_p^*+g_0$ with $A_0\in(0,\epsilon)$ 
and $\|g_0\|<A_0^2$,
taking $\epsilon< \sigma/(C_{d,p,q}+1)$ we guarantee that 
$f_1=R_{L,0}f_0$ is well defined and from Lemma \ref{estimativas}, 
it follows that $f_1$ can be written as $f_1=A_1R^0_{L}f_p^*+g_1$ 
with $A_1$ and $g_1$ given respectively by (\ref{def:An}) and 
(\ref{def:g_j+1:cas:mar}) with $n=0$. 
From (\ref{cotaA:m}) with $n=0$, using that
$\|g_0\|<A_0^2<1$,
$|\lambda|<\mu<1$ and $\alpha\geq 2$, we get
$|A_1-A_0+\mu\beta_0A_0^{\alpha_c}|\leq 7\Lambda \mu A_0^{\alpha_c+1}$,
or
\begin{equation}
\label{des:A1}
A_0[1-\mu A_0^{\alpha_c-1}(\beta_0+7\Lambda A_0)]<A_1<A_0
[1+\mu A_0^{\alpha_c-1}(-\beta_0+7\Lambda A_0)].
\end{equation}
Notice that, since $1>A_0>0$, if $\mu[\beta_0+7\Lambda]<1$, then 
$A_1>0$ from the left hand side of the inequality above and,
from the right hand side, for small $A_0$, that is, if 
$7\Lambda A_0<\beta_0$, we get $A_1<A_0$.
It follows from (\ref{cotag}) with $n=0$ that
$$
\|g_1\|\leq \Big(\frac{C}{L^{(p+1)/d}}+8\Lambda\mu\Big)A_0^2.
$$
Since $A_0<1$, it follows from (\ref{des:A1}) that $A_1^2>A_0^2[1-\mu(7\Lambda+\beta_0)]^2$ and, therefore, if
\begin{equation}
\label{cota:lambda}
\frac{C}{L^{(p+1)/d}}+8\Lambda\mu<[1-\mu(7\Lambda+\beta_0)]^2,
\end{equation}
then $\|g_1\|<A_1^{2}$.
Inequality (\ref{cota:lambda}) is valid if we take
$$
\mu<\frac{1-CL^{-(p+1)/d}}{22\Lambda+2\beta_0}.
$$
Notice that if $L>L_2$, the right hand side of the above inequality is positive. Now define
\begin{equation}
\label{def:epsilon:cas:mar}
\epsilon\equiv\min\{1,\beta_*/(7\Lambda),\sigma/(C_{d,p,q}+1), \sigma/(\tilde K+1)\}
\end{equation}
and
\begin{equation}
\label{def:lambda:cas:mar}
\lambda^*\equiv\min\left\{\gamma, \frac{1}{7\Lambda+\beta^*}, \frac{1-CL^{-(p+1)/d}}{22\Lambda+2\beta^*}\right\},
\end{equation}
where 
$\gamma$ and $\Lambda$ are given by the Renormalization Lemma. 

Suppose $0<A_n<A_{n-1}<\epsilon$, $\|g_{n-1}\|<A_{n-1}^2$ for $n=1,\cdots,k$ and $\lambda<\lambda^*$.
We shall prove that $0<A_{k+1}<A_{k}$ and $\|g_{k}\|<A_{k}^2$.
Taking $\epsilon<\sigma/(\tilde K+1)$, it is easy to see from (\ref{lem:pon:fix}) that
$f_{k+1}=R_{L,k}f_k$ is well defined and it can be decomposed as in (\ref{rep:fn}).
From (\ref{cotaA:m}), and using the induction hypothesis, we get
\begin{equation}
\label{des:Ak}
A_k[1-\mu A_k^{\alpha_c-1}(\beta_k+7\Lambda A_k)]<A_{k+1}<A_k[1+\mu A_k^{\alpha_c-1}(-\beta_k+7\Lambda A_k)].
\end{equation}
Since $0<\mu <\lambda^*$, $\beta_k\leq\beta^*$ and $0<A_k<A_{k-1}<...<A_0<\epsilon$, it follows from the left hand side of (\ref{des:Ak}) that $A_{k+1}>0$ and from the right hand side of (\ref{des:Ak}) that $A_{k+1}<A_k$. To show that $\|g_{k+1}\|<A_{k+1}^2$, we take (\ref{cotag}) with $n=k$ and use the induction hypothesis $\|g_{k}\|<A_{k}^2<1$ and $\alpha_c\geq 2$ to get
$$
\|g_{k+1}\|\leq \Big(\frac{C}{L^{(p+1)/d}}+8\Lambda\mu\Big)A_k^2.
$$
Once again, since $A_k<1$, it follows from (\ref{des:Ak}) that
 $A_{k+1}^2>A_{k}^2[1-\mu(7\Lambda+\beta_k)]^2$, and therefore, to prove that $\|g_{k+1}\|<A_{k+1}^{2}$, we need that
$$
\frac{C}{L^{(p+1)/d}}+8\Lambda\lambda<[1-\mu(7\Lambda+\beta_k)]^2,
$$
which is true since $[1-\mu(7K+\beta_k)]^2\geq 1-2\mu(7K+\beta_k)$, $\lambda<\lambda^*$, $\beta_k<\beta^*$ and $L>L_2$.

We have just shown that there exists $A=\lim_{n\rightarrow\infty} A_n$ 
and $0\leq A<\epsilon$. We will now prove that $A=0$. 
Taking the limit $k\rightarrow\infty$ in (\ref{des:Ak}), 
since $\beta_k\rightarrow\beta$, we have
$$
\mu A^{\alpha_c-1}(\beta-7\Lambda A)\leq 0.
$$
Since $A<\epsilon<\beta/(7\Lambda)$, 
it follows that $\beta-7\Lambda A>0$ and, since $\mu>0$, we have $A=0$.

\eop

We finally prove Theorem \ref{teo:pri:cas:mar}. We first prove that 
(\ref{eq:pri:cas:mar}) holds for the sequence $t=L^n$, with $L>L_2$, 
and then we extend this result.
\begin{lema}
\label{lema:princ}
Consider $L>L_2$ and suppose $\lambda^*$ and $\epsilon$ 
are given respectively by (\ref{def:epsilon:cas:mar}) 
and (\ref{def:lambda:cas:mar}). Suppose also that hypothesis 
$\textbf{(M)}$ are valid and that $0<A_0<\epsilon$, 
$0<\mu+|\lambda|<\lambda^*$ and $|\lambda|<\mu$. 
Then, the unique solution 
$u$ to (\ref{equ:int:cas:mar}) satisfies 
\begin{equation}
\label{lim:n}
\lim_{n\rightarrow\infty}
\|L^{n(p+1)/d}u(L^{n(p+1)/d}.,L^n)-[\mu(\alpha_c-1)\beta n]^{-(p+1)/d}f_p^*\|=0.
\end{equation}
\end{lema}

\textbf{Proof: } From lemmas \ref{estimativas} and \ref{lem:4} 
and using that $\|g_n\|<A_n^{2}$  and $\|g_n\|\leq 1$ in (\ref{cotaA:m}), we get
$$
A_{n+1}=A_n-\mu\beta_nA_n^{\alpha_c}+O(A_n^{\alpha_c+1})~~~~~~n=0,1,2,\cdots.
$$
Therefore
$$
A_{n+1}^{\alpha_c-1}=A_n^{\alpha_c-1}[1-\mu\beta_nA_n^{\alpha_c-1}+O(A_n^{\alpha_c})]^{\alpha_c-1}=
A_n^{\alpha_c-1}[1-\mu\beta_n(\alpha_c-1)A_n^{\alpha_c-1}+O(A_n^{\alpha_c})].
$$
Defining $A_n=\nu_n^{-1}$,
$$
\nu_{n+1}^{\alpha_c-1}=\nu_n^{\alpha_c-1}[1-\mu(\alpha_c-1)\beta_n A_n^{\alpha_c-1}+O(A_n^{\alpha_c})]^{-1}.
$$
Since $\alpha_c\geq 2$ and $\lim_{n\rightarrow\infty}A_n=0$, for $n$ large enough
$|\mu(\alpha_c-1)\beta_n A_n^{\alpha_c-1}+O(A_n^{\alpha_c})|<1$ and
\begin{equation}
\label{equ:nu}
\nu_{n+1}^{\alpha_c-1}-\nu_n^{\alpha_c-1}=
\mu(\alpha_c-1)\beta_n+O(\nu_n^{-1}) ~~~(n\rightarrow\infty).
\end{equation}
It follows from (\ref{equ:nu}) that there is $n_0>0$ such 
that $\nu_{n+1}^{\alpha_c-1}-\nu_n^{\alpha_c-1}>\frac{\mu\beta_*(\alpha_c-1)}{2}$ for all $n>n_0$.
Therefore, for $n>2n_0$,
\begin{eqnarray*}
	\nu_n^{\alpha_c-1}&=&\nu_{n_0}^{\alpha_c-1}+\sum_{k=n_0}^{n-1}\nu_{k+1}^{\alpha_c-1}-\nu_k^{\alpha_c-1}>
	\nu_{n_0}^{\alpha_c-1}+\frac{\mu\beta_*(\alpha_c-1)(n-n_0)}{2} \\
	&>&
	\frac{\mu\beta_*n(\alpha_c-1)}{2}\Big(1-\frac{n_0}{n}\Big)>\frac{\mu\beta_*n(\alpha_c-1)}{4}
\end{eqnarray*}
and so $\nu_n^{-1}=O(n^{\frac{-1}{\alpha_c-1}})$. Using this in (\ref{equ:nu}) we get
$$
\nu_{n+1}^{\alpha_c-1}-\nu_n^{\alpha_c-1}=
\mu(\alpha_c-1)(\beta_n-\beta)+\mu(\alpha_c-1)\beta+O(n^{\frac{-1}{\alpha_c-1}}) ~~~(n\rightarrow\infty).
$$
From Lemma \ref{lem:cond:beta} 
we can write
$$
\nu_{n+1}^{\alpha_c-1}-\nu_n^{\alpha_c-1}=\mu(\alpha_c-1)\beta+O\left(n^{\frac{-1}{\alpha_c-1}}\right) ~~~(n\rightarrow\infty),
$$
and $\nu_{n}^{\alpha_c-1}=\mu(\alpha_c-1)\beta n+O\left(n^{\frac{\alpha_c-2}{\alpha_c-1}}\right)$( for $\alpha_c=2, \mbox{we have } O(\ln n) $). 
Therefore,
$$
A_n^{\alpha_c-1}=\left\{\mu(\alpha_c-1)\beta n\left[1+O\left(n^{\frac{-1}{\alpha_c-1}}\right)\right]\right\}^{-1}.
$$
Recalling that $\alpha_c=1+d/(p+1)$, we obtain
\begin{equation}
\label{cota:An:cas:marginal}
A_n=\left[\frac{1}{\mu (\alpha_c-1)\beta n}\right]^{(p+1)/d}+O\left(n^{-2/(\alpha_c-1)}\right).
\end{equation}
We use (\ref{cota:An:cas:marginal}) to get (\ref{lim:n}). Notice that
		$$
		\left\|L^{\frac{n(p+1)}{d}}u\left(L^{\frac{n(p+1)}{d}}.,L^n\right)-A_nR^0_{L^n}f^*_p \right\|+
		$$
		$$\frac{1}{[\mu(\alpha_c-1)\beta n]^{\frac{p+1}{d}}}\|R^0_{L^n}f_p^*-f_p^*\|+
		\left|A_n-\frac{1}{[\mu(\alpha_c-1)\beta n]^{\frac{p+1}{d}}}\right|\|R^0_{L^n}f^*_p \|
		$$
is an upper bound for $\|L^{n(p+1)/d} u(L^{n(p+1)/d}.,L^n)-[\mu(\alpha_c-1)\beta n]^{-(p+1)/d}f_p^*\|$. Then, since 
$f_n(x)=L^{n(p+1)/d}u(L^{n(p+1)/d}x,L^n)$, it follows from (\ref{lem:pon:fix}), (\ref{rep:fn}) and (\ref{cota:An:cas:marginal}), that the above bound is, 
for large $n$,  
\begin{equation}
\label{cota:lim:prin}
\left[\frac{1}{\mu (\alpha_c-1)\beta n}\right]^{\frac{2(p+1)}{d}}+
\frac{M}{[\mu(\alpha_c-1)\beta n]^{\frac{p+1}{d}}}\left|\frac{r(L^n)}{L^{n(p+1)}}
\right|^{\frac{1}{d}}+O\left(\frac{1}{n^{2/(\alpha_c-1)}}\right).
\end{equation}
Taking the limit $n\rightarrow\infty$, we get (\ref{lim:n}).

\eop

\textbf{Proof of Theorem \ref{teo:pri:cas:mar}: } We have proved that (\ref{eq:pri:cas:mar}) holds for
small $f$ and $t=L^n$ $(n=1,2,\cdots)$, for $L>L_2$. Recalling $\beta$ given by (\ref{def:beta}) and defining
$$
A\equiv\left\{\mu(\alpha_c-1)R\left[\frac{p+1}{(2\pi)^{d}}\right]^{\frac{1}{p+1}}\right\}^{-(p+1)/d}
$$
it follows from (\ref{cota:lim:prin}) that if $t=L^n$, then $\|t^{(p+1)/d}u(t^{(p+1)/d}\cdot,t)-A(\ln t)^{-(p+1)/d}f_p^*\|$ is bounded by
$$
\left[\frac{A}{(\ln t)^{1/(\alpha_c -1)}}\right]^{2}
+\frac{MA}{(\ln t)^{1/(\alpha_c-1)}}\left|\frac{r(t)}{t^{p+1}}
\right|^{\frac{1}{d}}+O\left(\left(\frac{\ln L}{\ln t}\right)^{2/(\alpha_c-1)}\right),
$$
where $M$ is the constant in (\ref{lem:pon:fix}). 
The result is obtained by extending the above bound as done in the proof of 
Theorem II.1 in \cite{bib:braga-mor-souza}.

\eop


\end{document}